\documentclass[aps,prd,twocolumn,border=10pt,superscriptaddress]{revtex4-2}
\usepackage{tikz}
\usetikzlibrary{decorations.pathmorphing}
\usetikzlibrary{shapes, arrows, positioning, patterns, calc}
\usepackage{epsfig}
\usepackage{amsmath,amssymb,amsfonts}
\usepackage{amsthm}
\usepackage{dsfont}
\usepackage{multirow}
\usepackage{appendix}
\usepackage{slashed}
\usepackage[active]{srcltx}
\usepackage{psfrag}
\usepackage[colorlinks,
linkcolor = red,
urlcolor  = blue,
citecolor = blue,
anchorcolor = blue]{hyperref}
\usepackage{booktabs}
\usepackage[useregional]{datetime2}
\DTMlangsetup[en-US]{ord=omit}
\usepackage{graphicx}

\usepackage{bbold}
\usepackage{dcolumn}

\usepackage{mathtools}
\usepackage{setspace}
\begin{document}

\title{$\chi_{c1}(4010)$ as a $D^{*\pm}D^{\mp}$ Molecular State: Mass, Decay Constant, and Width}

		\author{ S. Damen}
	\affiliation{Department of Physics, Kocaeli University, 41001 Izmit, T\"urkiye}
	
\author{H.~Sundu}

\affiliation{Department of Physics Engineering, Istanbul Medeniyet University, 34700
Istanbul, T\"urkiye}
\author{J.Y.~S\"ung\"u}
	\affiliation{Department of Physics, Kocaeli University, 41001 Izmit, T\"urkiye}
	
\begin{abstract}

The charmonium-like state $\chi_{c1}(4010)$, recently observed by the LHCb 
Collaboration in $B^+ \to D^{*\pm}D^{\mp}K^+$ decays, is investigated as an 
axial-vector $D^{*\pm}D^{\mp}$ molecular state within the QCD sum rule 
approach. A two-point sum rule incorporating quark, gluon, and mixed vacuum 
condensates up to dimension $10$ yields the mass and decay constant, 
$m_{\chi_{c1}(4010)} = (4005.89 \pm 47.15)~\text{MeV}$ and 
$f_{\chi_{c1}(4010)} = (2.87 \pm 0.55)\times 10^{-3}~\text{GeV}^{4}$, in 
agreement with the LHCb measurement. Since this mass lies above the 
$D^{*\pm}D^{\mp}$ threshold, the strong coupling constants at the 
$\chi_{c1}(4010)D^{*\pm}D^{\mp}$ vertices are extracted from three-point 
sum rules and used to compute the partial widths of the $D^{*-}D^{+}$ 
and $D^{*+}D^{-}$ channels. The resulting total width, 
$\Gamma_{\chi_{c1}(4010)} = (57.64 \pm 9.64)~\text{MeV}$, is compatible 
with experiment within uncertainties. These results support a 
$D^{*\pm}D^{\mp}$ molecular interpretation of $\chi_{c1}(4010)$ and offer 
sum-rule predictions for comparison with future measurements at BESIII, 
LHCb, and Belle~II.

\end{abstract}
\maketitle
\section{Introduction}\label{sec:Int}
States with quantum numbers or mass spectra that fall outside the standard classification of mesons $(q\bar{q})$ and baryons $(qqq)$, including multiquark configurations and gluon hybrids, have become a central topic in hadron spectroscopy. In the charmonium $(c\bar{c})$ and bottomonium $(b\bar{b})$ regions, several narrow resonances fail to match constituent quark model predictions for conventional heavy mesons \cite{Brambilla:2019esw, Lebed:2016hpi, Agaev:2020zad, Agaev:2016srl, Agaev:2018vag, Agaev:2021vur, Agaev:2026kyz, Agaev:2025nkw}. Their structure remains unresolved and is an open problem in nonperturbative QCD. Proposed interpretations include multiquark systems, hadronic molecules, and hybrid states \cite{Guo:2017jvc, Chen:2016qju, Esmer:2026jlb}. Several theoretical models have also been extended to directly accommodate these states.

The charmonium-like sector with $J^{PC}=1^{++}$ illustrates this problem clearly. 
The $\chi_{c1}(3872)$, usually treated as the first observed state in this channel, has been debated since its discovery~\cite{Belle:2003nnu, Wang:2023ovj, Belle:2008fma, Sundu:2016oda}. 
Its mass lies almost exactly at the $D^{0}\bar{D}^{*0}$ threshold, complicating a simple $\chi_{c1}(2P)$ assignment, and its decay width is far narrower than decay models predict for that state. The higher mass $\chi_{c1}(4140)$ shows the opposite pattern: both its mass 
and decay width are higher than a $\chi_{c1}(2P)$ charmonium assignment 
predicts~\cite{Shi:2024llv, Gao:2025tob, Agaev:2017foq, Agaev:2023gti, Agaev:2022pis}. 

In 2024, an exotic candidate with $J^{PC}=1^{++}$, denoted $\chi_{c1}(4010)$, was identified by the LHCb collaboration via  analysis of the $D^{*+}D^{-}$ invariant mass distribution in $B^{+}\to D^{*\pm}D^{\mp}K^{+}$ decays~\cite{LHCb:2024vfz}. The measured resonance parameters are
\begin{align}
m_{\chi_{c1}(4010)} &= (4012.5^{+3.6+4.1}_{-3.9-3.7})~\mathrm{MeV}, \notag \\
\Gamma_{\chi_{c1}(4010)} &= (62.7^{+7.0+6.4}_{-6.4-6.6})~\mathrm{MeV}. \label{eq:Data_chi}
\end{align}

Ding et al.~included $\chi_{c1}(4010)$ as an extra resonance in their amplitude analysis of $B^+ \to D^{*+} D^- K^+$~\cite{Ding:2025uhh}, motivated by the fact that molecular contributions from $\chi_{c1}(3872)$ and $T^{*}_{\bar{c}\bar{s}0}(2870)^0$ alone could not reproduce the LHCb spectrum. 
Adding $\chi_{c1}(4010)$ accounts for the excess near $4010  ~\mathrm{MeV}$ and brings the amplitude model into agreement with the data. The analysis uses $C$-parity relations to constrain charmonium(-like) contributions across the two charge-conjugate decay channels. 
Four such states decaying to $D^{*\pm}D^{\mp}$ are observed: $\eta_c(3945)$ ($0^{-+}$), $h_c(4000)$ ($1^{+-}$), $\chi_{c1}(4010)$ ($1^{++}$), and $h_c(4300)$ ($1^{+-}$). The amplitude model for $B^{+} \to D^{*+} D^{-} K^{+}$ also includes the exotic resonances $T^{*}_{\bar{c}\bar{s}0}(2870)^0$ and $T_{c\bar{s}1}^{*0}(2900)$, first seen in $B^{+} \to D^{+} D^{-} K^{+}$ decays, and confirms their presence in this channel as well. 
If the $T_{c\bar{s}0,1}^{*0}$ states are eventually confirmed, this cross-channel difference could help clarify how they are produced~\cite{LHCb:2024vfz}.
These results establish the $D^{*\pm}D^{\mp}$ threshold region as populated by multiple resonances, motivating a dedicated structural study of $\chi_{c1}(4010)$
and others.

The $1^{++}$ charmonium-like sector already includes 
$\chi_{c1}(4140)$, $\chi_{c1}(4274)$, and 
$\chi_{c1}(4685)$~\cite{LHCb:2020qdy, Olsen:2017bmm}, while 
$\chi_{c1}(3872)$ and $\chi_{c2}(3930)$ lie considerably 
lower~\cite{Dai:2026fkg}. In mass, $\chi_{c1}(4010)$~\cite{LHCb:2024vfz} 
sits between these two sets: above both of the latter and below 
$\chi_{c1}(4140)$ and $\chi_{c1}(4274)$. Its quantum numbers, $J^{PC}=1^{++}$, are what a conventional $\chi_{c1}$ charmonium state would have, but its mass does not match any established $\chi_{c1}$ state predicted by standard charmonium models~\cite{LHCb:2024vfz}. Such considerations have opened the door to several alternative interpretations, including a compact tetraquark \cite{Wang:2025sic}, a meson-meson molecule, or even a hybrid configuration that mixes conventional charmonium with exotic components \cite{Dai:2026fkg}. QCD sum rules (QCDSR) have similarly been applied to 
compute the mass spectrum of the heavy-quark spin symmetry partners of 
$X(3872)$ within a molecular framework~\cite{Mutuk:2018zxs}.

The $\chi_{c1}(4010)$ resonance plays a key role in coupled-channel analyses of the $\chi_{c1}(2P)$ sector. 
Its closeness to open-charm thresholds makes it sensitive to both the underlying $c\bar{c}$ core and hadronic molecular components, and it serves as a reference point for the strength of $c\bar{c}$-molecule mixing in hidden-charm states~\cite{Miyake:2025ktz, Miyake:2025mmp, Wang:2024ytk, Li:2024pfg}. Its position above open-charm thresholds constrains the properties of the bare $\chi_{c1}(2P)$ core and its coupling strength to $D\bar{D}^{*}$ channels. 
These analyses find that $X(3872)$ does not arise from a simple mass shift of $\chi_{c1}(2P)$, but from coupled-channel dynamics involving both charmonium and molecular components, with the predicted mass of $\chi_{c1}(4010)$ acting as a key indicator of this mixing. Lattice determinations of the $2\,{}^3P_1$ charmonium mass cluster around $4010\text{--}4067\,\mathrm{MeV}$~\cite{CP-PACS:2001yuj,Chen:2000ej}, so a conventional $\chi_{c1}(2P)$ assignment for $\chi_{c1}(4010)$ cannot be excluded on mass grounds alone. Our sum rule analysis shows that a $D^{*\pm} D^{\mp}$ molecular current also reproduces the observed mass and width, and the coupled-channel studies of Refs.~\cite{Shi:2024llv, Gao:2025tob} suggest that the physical state carries both components.

Coupling the bare $\chi_{c1}(2P)$ state to the $D\bar{D}^*$ channels produces two distinct poles in the complex energy plane: one is a bound state near the $D\bar{D}^*$ threshold, and the other originates from the bare state and shifts into the complex continuum~\cite{Deng:2023mza}. 
This structure yields a $D\bar{D}^{*}$ lineshape with a threshold enhancement and a broad resonance-like feature around $3990$~MeV, consistent with the analysis in Ref.~\cite{Gao:2025tob}. 
The two features correspond to the $\chi_{c1}(3872)$ and $\chi_{c1}(4010)$ states, respectively~\cite{Gao:2025tob}.

In the present investigation, we examine the resonance $\chi_{c1}(4010) $  as an axial vector configuration represented as $D^{*\pm}D^{\mp}$, characterized by the spin-parity $J^{PC} = 1^{++}$, and proceed to compute both the mass, decay constant, and the full decay width of this particular state. The analysis is carried out within the framework of QCDSR, which is a powerful nonperturbative approach in high-energy physics. We derive the mass and decay constant of $\chi_{c1}(4010)$ from the two-point sum rule. As shown later, its mass is large enough that the dominant decay channels are $D^{*+}D^{-}$ and $D^{*-}D^{+}$; we compute the corresponding partial widths using the three-point sum rule technique. For the decay $\chi_{c1}(4010) \to D^{*\pm} D^{\mp}$, this yields a sum rule for the strong coupling constant $g_{1,2}(q^2)$ at the $\chi_{c1}(4010)D^{*\pm}D^{\mp}$ vertex. Evaluating $g_{1,2}(q^2)$ at the $D^{\pm}$ mass shell, $q^2=m^2_{D^{\pm}}$, fixes $g_{1,2}$, which is then used to compute the partial decay width.

The paper is organized as follows. Sec.~\ref{sec:MassQCDSUMRULES} computes the mass and decay constant of the axial-vector molecular state $\chi_{c1}(4010)$. Sec.~\ref{sec:DecayQCDSUMRULES} analyzes the decay channels $\chi_{c1}(4010) \to D^{*\pm} D^{\mp}$ and determines the total decay width. Sec.~\ref{sec:Conclusions} presents our conclusions.
\section{QCDSR Formalism: THE SPECTROSCOPIC PARAMETERS OF THE MOLECULE $\chi_{c1}(4010)$}\label{sec:MassQCDSUMRULES}
The QCDSR method rests on the operator product expansion (OPE)~\cite{Shifman:1978bx, Shifman:1978by, Reinders:1984sr}, 
in which perturbative QCD at short distances expands a correlator into a 
power series in the squared momentum, with coefficients given by the 
Wilson coefficients. 
A Borel transformation improves the convergence of this series at low momentum, or equivalently long distance. 
The same correlator can be computed in two different ways: at the quark 
level, which yields the OPE (or QCD) side, and at the hadronic level, 
which produces the phenomenological side. A dispersion relation connects 
the two, and this connection defines the QCDSR, from which hadronic 
properties are extracted.

The mass and decay constant of $\chi_{c1}$ follow from the two-point QCDSR approach~\cite{Albuquerque:2018jkn, Agaev:2020zad}, which works well for both ordinary particles and multiquark hadrons. 
The central object in this framework is the hadron's interpolating current. 
For $\chi_{c1}$, the current $J_{\mu}(x)$ is constructed as~\cite{Chen:2021kol}%
\begin{eqnarray}
 J_{\mu}(x)&=&\frac {i}{\sqrt{2}}\Big(  \bar{c}_{a}(x) \gamma_{\mu} d_{a}(x) \, \bar{d}_{b}(x) \gamma_{5} c_{b}(x) \nonumber\\ &-& \bar{d}_{a}(x) \gamma_{\mu} c_{a}(x) \, \bar{c}_{b}(x) \gamma_5 d_{b}(x)  \Big),
\end{eqnarray}%
where $a,b$ are color indices,  $d$ and $c$ are the down and charm quark fields. It describes a molecular state with $J^{PC}=1^{++}$.

To obtain the sum rules for the mass and decay constant of $\chi_{c1}$, it is necessary to start by calculating the following correlation function~\cite{Reinders:1984sr}:
\begin{equation}\label{corr0_1}
\Pi_{\mu\nu} (p)= i \int d^4x e^{i p\cdot x} \langle 0 \vert  {\cal T} (J_\mu(x) {J_\nu^{\dagger}(0))} \vert 0 \rangle.
\end{equation}%
On the phenomenological side, $\Pi_{\mu\nu}(p)$ can be expressed in terms of the spectral parameters of $\chi_{c1}$. 
Inserting the full set of states with $J^{PC}=1^{++}$ into $\Pi_{\mu\nu}(p)$ and integrating over $x$ gives $\Pi_{\mu\nu}^{\mathrm{Phys}}(p)$:
\begin{align}\label{corr_phen_1x1}
\Pi_{\mu\nu}^{\mathrm{Phys}}(p)& =  
\frac{\langle 0|J_{\mu}|\chi_{c1}(p,\varepsilon)\rangle \langle \chi_{c1}(p,\varepsilon)|J_{\nu}^{\dag }|0\rangle }{m_{ \chi_{c1}}^{2}-p^{2}}  \nonumber\\
& + \dots\ .
\end{align}%

This is the hadronic representation of the correlator, $\Pi_{\mu\nu}^{\mathrm{Phys}}(p)$, the phenomenological side of the sum rule. 
The term written explicitly in Eq.~(\ref{corr_phen_1x1}) is the ground-state contribution from $\chi_{c1}$; the ellipsis represents contributions from radially excited states and the continuum. 
$\Pi_{\mu\nu}^{\mathrm{Phys}}(p)$ can be reformulated in a more convenient form by means of the matrix element
\begin{equation}\label{corr2a}
 \langle 0|J_{\mu}|\chi_{c1}(p,\varepsilon)\rangle =  m_{\chi_{c1}} f_{\chi_{c1}}  \, \varepsilon_{\mu}(p),
\end{equation}
where $\varepsilon _{\mu }$ is the polarization vector of the axial-vector state. Accordingly, the expression  $\Pi_{\mu\nu}^{\mathrm{Phys}}(p)$ in terms of the parameters $m$ and $f$  is easily expressed as: 
\begin{equation}\label{phenS}
\Pi_{\mu\nu}^{\rm Phys}(p)=\frac{m^2_{\chi_{c1}} f^2_{\chi_{c1}}} 
{m_{\chi_{c1}}^2-p^2} \Big(g_{\mu\nu}-\frac {p_{\mu}p_{\nu}}{m_{\chi_{c1}}^2}\Big)~  +...\ .
\end{equation}
Two independent Lorentz structures parameterize the correlation function. We use the $\sim g_{\mu\nu}$ term for the sum rule analysis 
since it only receives a contribution from the spin-$1$ particle. This 
leads to the invariant amplitude
\begin{equation}
\Pi^{\rm Phys}(p) = \frac{m^2_{\chi_{c1}} f^2_{\chi_{c1}}}
{m_{\chi_{c1}}^2-p^2},
\end{equation}
which is written as the r.h.s. of Eq.~(\ref{phenS}).

The QCD side is obtained by substituting the explicit form of $J_{\mu}(x)$ 
into Eq.~(\ref{corr0_1}) and contracting the heavy and light quark fields, 
which expresses the result in terms of the charm- and light-quark 
propagators:

\begin{widetext}
\begin{eqnarray}
\Pi_{\mu\nu}^{\mathrm{OPE}}(p) &=& \frac {1}{2} \int d^4x\, e^{ip \cdot x} \Big\{ 
\mathrm{Tr} \left[ \gamma_{\mu}\, S_d^{aa'}(x)\, \gamma_{\nu}\, S_c^{a'a}(-x) \right] 
\mathrm{Tr} \left[ \gamma_5\, S_c^{bb'}(x)\, \gamma_5\, S_d^{b'b}(-x) \right] \notag \\
& - & \mathrm{Tr} \left[ \gamma_{\mu}\, S_d^{ab'}(x)\, \gamma_5\, S_c^{b'a}(-x) \right]
\mathrm{Tr} \left[ \gamma_5\, S_c^{ba'}(x)\, \gamma_{\nu}\, S_d^{a'b}(-x) \right] \notag \\
& - & \mathrm{Tr} \left[ \gamma_\mu\, S_c^{ab'}(x)\, \gamma_5\, S_d^{b'a}(-x) \right]
\mathrm{Tr} \left[ \gamma_5\, S_d^{ba'}(x)\, \gamma_\nu\, S_c^{a'b}(-x) \right] \notag \\
& + & \mathrm{Tr} \left[ \gamma_\mu\, S_c^{aa'}(x)\, \gamma_\nu\, S_d^{a'a}(-x) \right]
\mathrm{Tr} \left[ \gamma_5\, S_d^{bb'}(x)\, \gamma_5\, S_c^{b'b}(-x) \right] \Big\}.\label{eq:CorrM_2}
\end{eqnarray}

\end{widetext}%
For the light d quark, we use the propagator in coordinate space, which takes the form 
 \begin{eqnarray}\label{lightquarkpropagator}
S_{d}^{ij}(x)&=& i\frac{\!\not\!{x}}{
2\pi^2 x^4}\delta_{ij}
-\frac{m_d}{4\pi^2 x^2}\delta_{ij}-\frac{\langle
\bar{d}d\rangle}{12}\Big[1-i\frac{m_d}{4} \!\not\!{x} \Big]\delta_{ij} \nonumber\\
&-&\frac{x^{2}}{192} m_{0}^{2}
\langle
\bar{d}d\rangle\Big[1-i\frac{m_d}{6} \!\not\!{x} \Big]\delta_{ij}
\nonumber\\
&-&\frac{ig_s G^{\alpha\beta}_{i j}}{32\pi^{2} x^{2}}
\Big(\!\not\!{x} \sigma^{\alpha\beta}+\sigma^{\alpha\beta} \!\not\!{x} \Big)\delta_{ij}-i\frac{x^2 \!\not\!{x} g_{s}^{2} \langle
\bar{d}d\rangle^{2}}{7776}\delta_{ij}\nonumber\\
&-&\frac{x^4 \langle
\bar{d}d\rangle \langle
g_{s}^{2} G^2\rangle}{27648} + \cdots .
\end{eqnarray}

The heavy quark propagator \( S_c^{ij}(x) \)  reads:%
\begin{eqnarray}\label{heavypropagator}
S_c^{ij}(x)&=&i \int\frac{d^4k\, e^{-ik \cdot x}}{(2\pi)^4}
\Bigg\{ \frac{\!\not\!{k}+m_c}{k^2-m_c^2}\,\delta_{ij}\nonumber\\
&-&\frac{g_{s}G^{\alpha\beta}_{ij}}{4}\,
\frac{\sigma^{\alpha\beta}(\!\not\!{k}+m_c)+(\!\not\!{k}+m_c)\sigma^{\alpha\beta}}{(k^2-m_c^2)^{2}}
\nonumber\\
&+&\frac{g_{s}^{2}G^2}{12}\,\delta_{ij}\,m_{c}\,
\frac{k^{2}+m_{c}\!\not\!{k}}{(k^{2}-m_{c}^{2})^{4}}
\nonumber\\
&+&\frac{g_{s}^{3}G^{3}}{48}\,\delta_{ij}\,
\frac{(\!\not\!{k}+m_{c})}{(k^{2}-m_{c}^{2})^{6}}
\Big[\!\not\!{k}\,(k^{2}-3m_{c}^{2})
\nonumber\\
&+&2m_{c}\,(2k^{2}-m_{c}^{2})\Big]
(\!\not\!{k}+m_{c})+\cdots\Bigg\}.
\end{eqnarray}

The correlation function $\Pi_{\mu\nu}^{\mathrm{OPE}}(p)$ is computed from 
the quark propagators to a fixed order in the OPE. It has a Lorentz 
structure proportional to the term $g_{\mu\nu}$, with invariant amplitude 
$\Pi^{\mathrm{OPE}}(p)$. Equating $\Pi^{\mathrm{OPE}}(p)$ to the 
phenomenological amplitude $\Pi^{\mathrm{Phys}}(p)$ gives the sum rule, on 
the phenomenological side of which the ground-state contribution appears 
together with those of higher resonances and the continuum; these latter 
contributions must be separated from the ground state. We first apply 
the Borel transformation to both sides, which suppresses the 
higher-resonance contributions exponentially and introduces explicit 
dependence on the Borel parameter $M^2$. The continuum is then removed 
via quark-hadron duality, which introduces the threshold parameter $s_0$. The sum rule is constructed by equating the OPE and phenomenological sides of the correlator~\cite{Colangelo:2000dp}:%
\begin{equation}
\mathcal{B}\Pi_{\mu\nu}^{\text{OPE}}(p^2) \equiv \mathcal{B}\Pi_{\mu\nu}^{\text{Phys}}(p^2).
\end{equation}
By applying the Borel transformation, the main contribution to  $\Pi^{\mathrm{Phys}}(p)$ takes a simple form%
\begin{equation}
\mathcal{B}\Pi ^{\mathrm{Phys}}(M^{2})=m^2_{\chi_{c1}} f^2_{\chi_{c1}} e^{-{\frac{m^{2}_{\chi_{c1}}}{M^{2}}}}.
\end{equation}%
Following the Borel transformation and continuum subtraction, the amplitude $\Pi_{\mu\nu} ^{\mathrm{OPE}}(p^{2})$  reads
\begin{equation}
\Pi (M^{2},s_{0})=\int_{4(m_{c}+m_{d})^{2}}^{s_{0}}ds\rho ^{\mathrm{OPE}%
}(s)e^{-s/M^{2}}+\Pi (M^{2}).  \label{eq:InvAmp}
\end{equation}

The first term in Eq.~(\ref{eq:InvAmp}) accounts for the bulk of the contributions and is written in terms of the two-point spectral density $\rho^{\mathrm{OPE}}(s)$, taken as the imaginary part of the correlator. The following term $\Pi (M^{2})$ includes nonperturbative contributions that are extracted directly from 
$\Pi_{\mu\nu}^{\mathrm{OPE}}(p)$.

The sum rules for $m$ and $f$ can be written as: %
\begin{equation}
m^{2}_{\chi_{c1}}=\frac{\Pi ^{\prime }(M^{2},s_{0})}{\Pi (M^{2},s_{0})}  \label{eq:Mass}
\end{equation}%
and%
\begin{equation}
f^{2}_{\chi_{c1}} =\frac{ e^{\frac{m^{2}_{\chi_{c1}}}{M^{2}}}}{
m^{2}_{\chi_{c1}}} \Pi(M^{2}, s_{0}), \label{eq:Coupling}
\end{equation}
where $ \Pi^{\prime}(M^{2},s_{0})=d\Pi(M^{2},s_{0})/d(-1/M^{2})$. The mass and decay constant sum rules given by Eqs. (\ref{eq:Mass}) and (\ref{eq:Coupling}) also contain quark, gluon, and mixed condensates that are considered as universal parameters and well-known objects in the computational frameworks \cite{Shifman:1978bx, Shifman:1978by, Ioffe:1981kw, Ioffe:2005ym, Narison:2015nxh}. The corresponding numerical values for all of these parameters are given below. We have also included the charm and light quark masses in Eq. (\ref{eq:Mass}). The nonperturbative contributions to $\Pi (M^{2},s_{0}) $ coming from vacuum condensates up to the dimension $10$ are contained in it. For the numerical calculations, we set $m_{d}^2=0$  and retain only the terms proportional to $m_d$.

The extracted mass and decay constant depend on two auxiliary parameters, the Borel mass $M^2$ and the continuum threshold $s_0$. 
Physical observables should not depend on these in principle, but the procedure for separating the ground-state contribution from higher resonances and the continuum introduces a residual dependence on both. 
Standard criteria are used to restrict $M^2$ and $s_0$ to a working window that satisfies two conditions: the pole contribution must dominate, and the OPE must converge. 
Both conditions are needed for the sum rule predictions to be trustworthy. 
The pole contribution, $\mathrm{PC}$, is defined by~\cite{Agaev:2019qqn}
\begin{equation}
\mathrm{PC}=\frac{\Pi (M^{2},s_{0})}{\Pi (M^{2},\infty )},  \label{eqPC}
\end{equation}%
whereas to check the OPE convergence, we employ
the ratio
\begin{equation}
R(M^{2})=\frac{\Pi ^{\mathrm{DimN}}(M^{2},s_{0})}{\Pi (M^{2},s_{0})},
\label{eq:Convergence}
\end{equation}
where $\Pi^{\mathrm{DimN}}(M^{2},s_{0})$ is the summation of the last few terms in the OPE, with $\mathrm{DimN}=\mathrm{Dim(8+9+10)}$.

The functions $R(M^{2})$ and $\mathrm{PC}$ help to establish the lower and upper limits for the Borel parameter, respectively. More in detail at $M_{\mathrm{min}}^{2}$ the $R(M_{\mathrm{min}}^{2})$ must be smaller than a certain prefixed number and at $M_{\mathrm{max}}^{2}$ the pole contribution $\mathrm{PC}$ must be larger than a minimal allowed value for this quantity. In the following, we apply the following requirements on $\mathrm{PC}$ and $R(M^{2})$ in the current analysis: %
\begin{equation}
\mathrm{PC}\geq 0.5,\ \ R(M_{\mathrm{min}}^{2})\leq  0.05 .  \label{eq:Const}
\end{equation}

\begin{table}[h!]
\centering
\caption{Input parameters.}
\label{tab:input_params}
\begin{tabular}{|l|l|}
\hline
\textbf{Parameter} & \textbf{Value} \cite{Narison:2015nxh,Ioffe:2005ym,ParticleDataGroup20245,Gubler:2018ctz,Narison:2010cg,Shifman:1978bx, Shifman:1978by,Yu:2026sso} \\ \hline

$m_d$ &  $4.67^{+0.48}_{-0.17}~\mathrm{MeV}$  \\ \hline
$m_c$ & $1.28 \pm 0.03~\mathrm{GeV}$  \\ \hline
$\langle \bar{d} d \rangle$ &  $(-0.24 \pm 0.01)^3~\mathrm{GeV}^3$   \\ \hline
$\left\langle \frac{\alpha_s}{\pi} G^2 \right\rangle$ & $(0.012 \pm 0.004)~\mathrm{GeV}^4$   \\ \hline
$\left\langle g_{s}^{3} G^3 \right\rangle$ & $(0.57\pm 0.29) \text{GeV}^{6}$ \\ \hline
$m_0^2$ & $0.8 \pm 0.1\ \text{GeV}^2$  \\ \hline
\end{tabular}
\end{table}%

Two conditions determine the allowed windows for $M^2$ and $s_0$: the perturbative part of the correlation function $\Pi(M^2, s_0)$ must dominate over the nonperturbative contributions, as shown in Fig.~\ref{fig:pert and nonper}, and the extracted hadronic properties must remain stable against variation of the Borel parameter $M^2$. 
Applying these criteria in the numerical analysis gives the following acceptable windows for $M^2$ and $s_0$:
\begin{equation}
M^{2}\in \lbrack 3,4]~\mathrm{GeV}^{2},\ s_{0}\in \lbrack 19.5,20.5]~\mathrm{GeV}^{2}.  \label{eq:Wind1}
\end{equation}%
These values satisfy the described restrictions. It can be seen that the pole contribution remains above $0.5$ for all $M^2$ and $s_0$ values considered according to Eq. (\ref{eq:Wind1}).

Predictions regarding the mass and the decay constant are derived by averaging the values computed across different selections of the Borel parameter $M^2$ and the continuum threshold $s_0$:
\begin{eqnarray}\label{eq:Result1}
m_{\chi_{c1}} &=&( 4005.89\pm 47.15)~\mathrm{MeV},  \notag \\ 
f_{\chi_{c1}} &=&(2.87\pm 0.55)\times 10^{-3}~\mathrm{GeV}^{4} .  
\end{eqnarray}
 \begin{figure}[h!]
     \centering
     \includegraphics[totalheight=6cm,width=8cm]{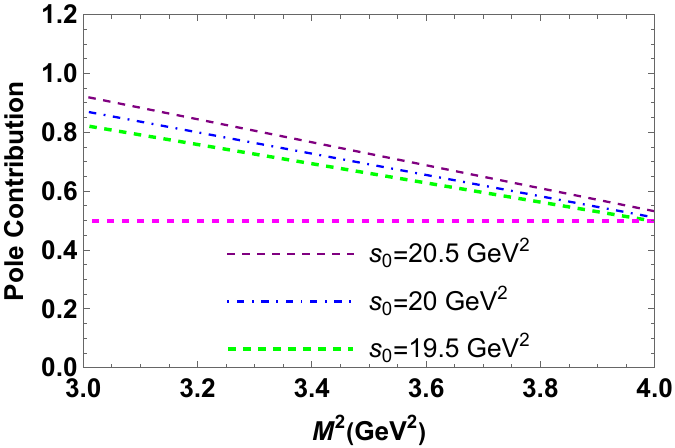}
     \caption{ Dependence of the pole contribution $(PC)$ on the Borel parameter $M^{2}$ for several values of $s_{0}$. The horizontal line corresponds to $\mathrm{PC}=0.5$, establishing the dominance condition.}
     \label{fig:PC with M2 at serious s0}
 \end{figure}%

Figs.~\ref{fig:Mass and borel and s0} and~\ref{fig:f and borel and s0} show the mass and decay constant of the $D^{*\pm}D^{\mp}$ molecule state as functions of the Borel parameter $M^2$ and the continuum threshold $s_0$. 
For the sum rule analysis to be reliable, the physical quantities 
should stay stable as the Borel parameter changes.
In practice, the extracted mass depends to some extent on the chosen working region: even within a Borel window that gives a quasi-stable prediction for $m$, some residual dependence remains, and this residual dependence is the main source of theoretical error in the sum rule predictions. 
The errors on the mass are smaller than those on the decay constant, because the mass comes from a ratio of correlators, which cancels part of the $M^2$ and $s_0$ dependence. 
The decay constant, by contrast, depends explicitly on the correlation function $\Pi(M^2,s_0)$, so it is more sensitive to changes in these auxiliary parameters. %

Therefore, the theoretical errors on the mass and decay constant are estimated to be around $\pm 1\%$ and $\pm 20\%$, respectively. The mass of the resonance $\chi_{c1}$ considered as an axial vector molecular state is consistent with the experimental data of the LHCb collaboration. Based on this correspondence, we conclude that $\chi_{c1}$ can be interpreted as a molecular state with quantum numbers $J^{PC}=1^{++}$.

The spectroscopic parameters of the molecular state $\chi_{c1}$ derived here will be used in the next section to estimate the width of its decay channels.%

\begin{widetext}

\begin{figure}[h!]
\begin{center}
\includegraphics[totalheight=6cm,width=8cm]{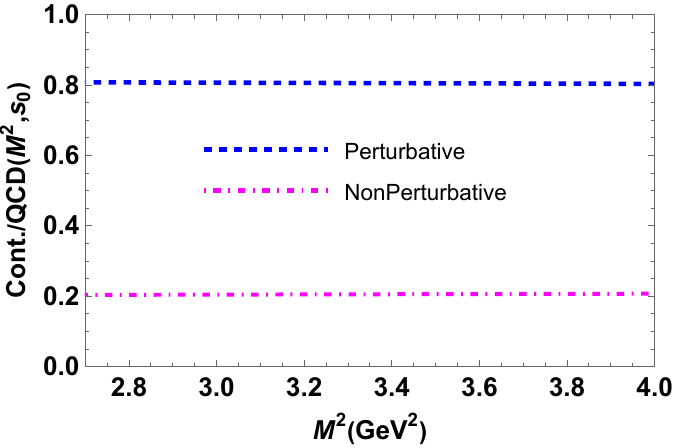}
\includegraphics[totalheight=6cm,width=8cm]{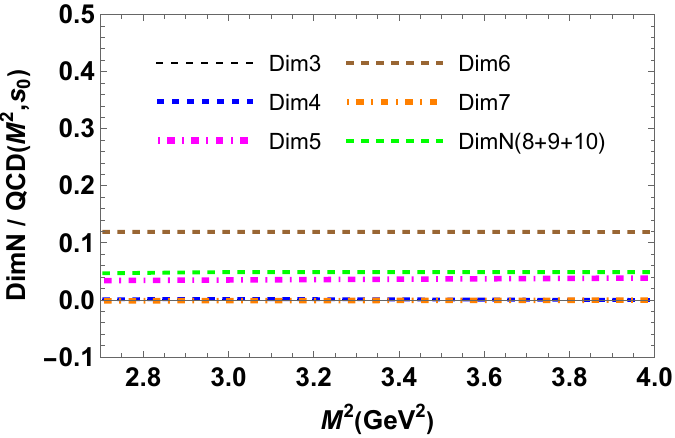}
\end{center}
\caption{Normalized contributions to $\Pi(M^{2}, s_{0})$ as functions of the Borel parameter. All curves are evaluated at $s_{0}=20~\mathrm{GeV}^{2}$.}
\label{fig:pert and nonper}
\end{figure}

\begin{figure}[h!]
\begin{center}
\includegraphics[totalheight=6cm,width=8cm]{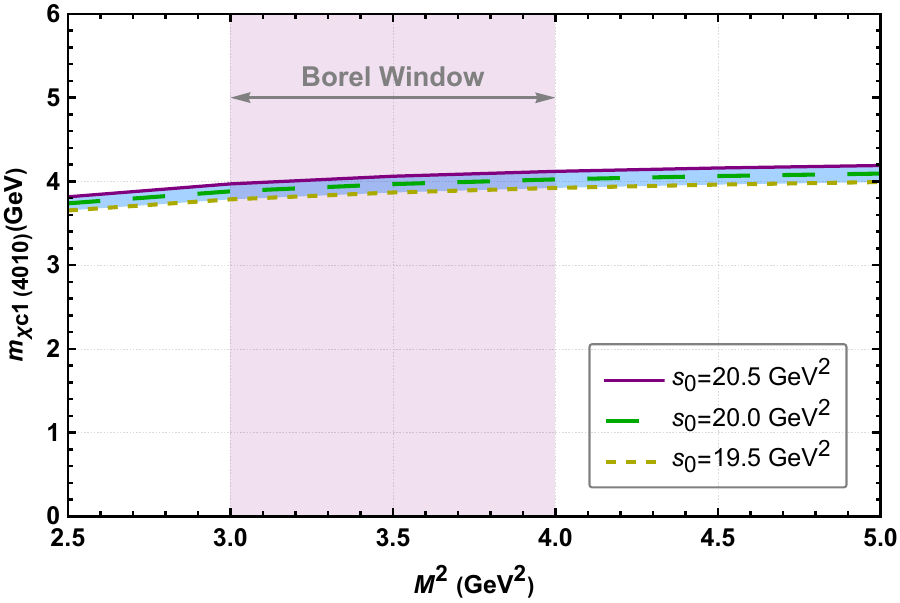}
\includegraphics[totalheight=6cm,width=8cm]{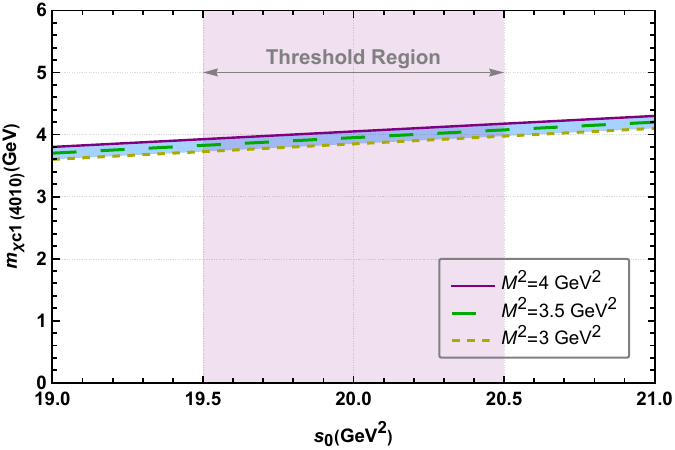}
\end{center}
\caption{
Variation of the molecular mass $m_{\chi_{c1}}$ with respect to the Borel parameter $M^{2}$ (left) and the continuum threshold $s_{0}$ (right) for selected parameter values.}
\label{fig:Mass and borel and s0}
\end{figure}

\begin{figure}[h!]
\begin{center}
\includegraphics[totalheight=6cm,width=8cm]{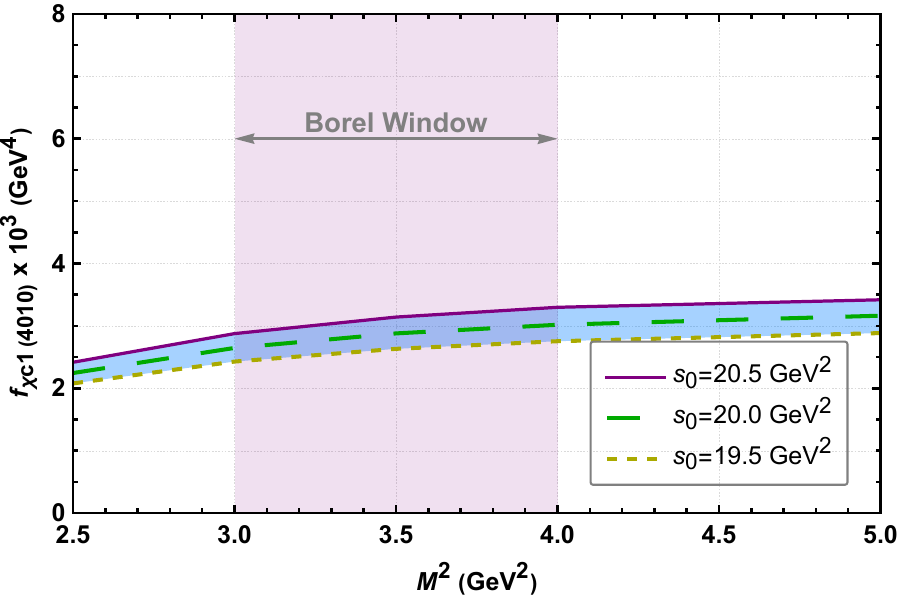}
\includegraphics[totalheight=6cm,width=8cm]{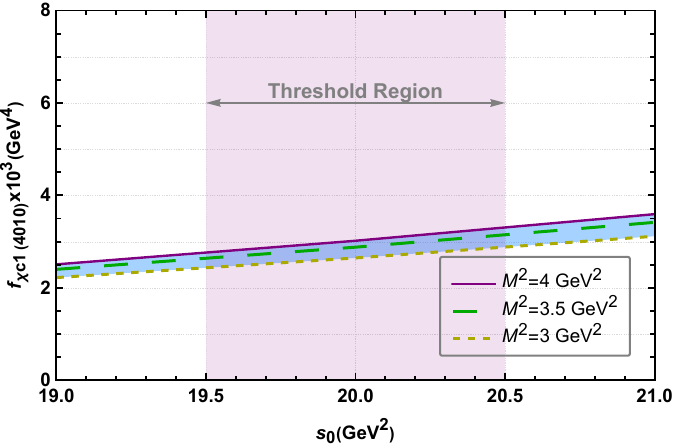}
\end{center}
\caption{The dependence of the molecular decay constant $f_{\chi_{c1}}$ on the Borel parameter $M^{2}$  (left panel) and the continuum threshold $s_{0}$ (right panel).
}
\label{fig:f and borel and s0}
\end{figure}

\end{widetext}
\section{STRONG COUPLINGS AND DECAY WIDTH OF $\chi_{c1}(4010)$}
\label{sec:DecayQCDSUMRULES}
Reliable values for the strong coupling constants are needed to understand strong interaction dynamics and hadronic processes in the nonperturbative regime of QCD. 
QCDSR is one of the most effective nonperturbative approaches for studying hadronic properties at low energies, and it has been applied to a range of mesonic vertices, including $D^{*+}D^-$ and $D^{*-}D^+$, to determine their strong couplings more precisely. 
These couplings enter directly into descriptions of hadronic decays and hadron-charm interactions, and they constrain the internal structure of exotic states and the mechanisms behind charm-flavored hadronic system formation. 

The spectroscopic properties of the molecular candidate $\chi_{c1}$ constrain its kinematically accessible decay modes. 
Since $\chi_{c1}$ has been observed in the $D^{*+}D^{-}$ and $D^{*-}D^{+}$ invariant-mass spectra, $B^{+} \to D^{*\pm} D^{\mp} K^{+}$ is likely its dominant production mechanism, with no alternative decay channels yet confirmed experimentally ~\cite{LHCb:2024vfz, Li:2024pfg}.

Here we study the $D^{*+}D^{-}$ and $D^{*-}D^{+}$ decay channels of the 
axial-vector molecule $\chi_{c1}(4010)$ and predict the corresponding 
partial decay widths.
These are fall-apart (OZI-superallowed) decays: all the valence quarks of the initial $\chi_{c1}$ go directly into the final-state mesons. 
By considering its mass and spin-parity quantum numbers $J^{PC}=1^{++}$ of $\chi_{c1}$, we identify its possible decay modes. 
The decay to the meson pair $D^{*\pm}D^{\mp}$ is an open channel for the transition of $\chi_{c1}$ into ordinary hadronic states.

The strong coupling constant $g$ governs the decay width of 
$\chi_{c1}(4010) \to D^{*+}D^{-}/D^{*-}D^{+}$. Apart from $g$, this width 
depends on the masses of the initial- and final-state mesons and their 
decay constants. The mass and decay constant of $\chi_{c1}(4010)$ were 
determined in the present work, while the parameters of the final-state 
mesons $D^{*\pm}$ and $D^{\pm}$ are taken from Table~\ref{tab:nput params2}. 
In this section, we carry out a detailed study of the strong decay 
processes into $D^{*+}D^{-}$ and $D^{*-}D^{+}$~\cite{Agaev:2025fwm}.
\subsection{$\chi_{c1} \rightarrow D^{*-} D^{+}$}

We initiate our examination with the first decay channel, $D^{*-} D^{+}$. The first step is to calculate the strong coupling constant related to the vertex $\chi_{c1} D^{*-} D^{+}$, which describes the strong interaction between the exotic molecular state and the two conventional mesons. To formulate  the strong coupling $g$, we employ the three-point QCDSR technique for this vertex, starting with an analysis of the following correlation function:%
\begin{align}
\Pi(p,p') &= i^2 \int d^4x \, d^4y \, e^{i(p'\cdot y - p \cdot x)} \nonumber \\
&\times \langle 0 | \mathcal{T} \{ J^{D^{*-}}(y) J^{D^{+}}(0) J^{\dagger}(x) \} | 0 \rangle. \label{eq:CF2}
\end{align}%
Here, $J(x)$ represents the interpolating current associated with the molecular state $\chi_{c1}$, whereas $J^{D^{*-}}(y)$ and $J^{D^+}(0)$ denote the currents corresponding to the vector and pseudoscalar mesons $D^{*-}$ and $D^{+}$, respectively. The four-momenta of $\chi_{c1}$ and $D^{*-}$ are represented by the symbols $p$ and $p'$, respectively, while the momentum of the $D^{+}$ meson is expressed as $q = p - p'$. The definitions of the mesonic currents are as follows:
\begin{align}
J_{\mu}^{D^{*-}}(x) &= \bar{c}_j(x)  i\gamma_{\mu} d_j(x),\\
J^{D^+}(x) &= \bar{d}_i(x)  i\gamma_5 c_i(x), \label{eq:interpolating current D positive negative1}
\end{align}
with $i$ and $j$ as color indices.

To derive the sum rule, by considering the physical manifestation of the correlation function $\Pi(p,p')$, which corresponds to the phenomenological side of the sum rule and is written as $\Pi^{\mathrm{Phys}}(p,p')$. This portion is mainly saturated by the contributions from the ground-state mesons, and hence we have:%
\begin{eqnarray}
\Pi^{\mathrm{Phys}}(p, p') &=& 
\frac{\langle 0 | J^{D^{*-}} | D^{*-}(p') \rangle 
\langle 0 | J^{D^+} | D^+(q) \rangle}{(p'^2 - m_{D^{*-}}^2)(q^2 - m_{D^+}^2)} \nonumber\\
& & \nonumber\\
& \times & \frac{\langle D^+(q) D^{*-}(p') | \chi_{c1}(p) \rangle 
\langle \chi_{c1}(p) | J^{\dagger} | 0 \rangle}{(p^2 - m_{\chi_{c1}}^2)} \nonumber\\
& + & \cdots .
\label{eq:CF33}
\end{eqnarray}%

In Eq.~\eqref{eq:CF33}, $m_{D^+}$ is the mass of the $D^+$ meson, and the ellipses denote contributions from higher resonances and continuum states. 
We simplify this expression using the relevant matrix elements, defined in terms of the physical parameters of the mesons. 
Inserting these matrix elements
\begin{equation}
\langle 0|J_{\mu }^{D^{*-}}|D^{*-} (p^{\prime },\varepsilon )\rangle
=f_{D^{*-} }m_{D^{*-} }\varepsilon _{\mu }(p^{\prime }),  \label{eq:Matrix D ngative }
\end{equation}
\begin{eqnarray}
\langle 0|J^{ D^{+}}| D^{+}(q)\rangle =\frac{f_ {D^{+}}m_{ D^{+}}^{2}}{m_{c}+m_{d}}. \label{eq:matrix eleman D positive}
\end{eqnarray}%

We describe the vertex $\chi_{c1} D^{*-} D^{+} $  by the matrix element%
\begin{align}
\langle D^{+}(q)D^{*-} (p^{\prime },\varepsilon )|\chi_{c1}(p,\epsilon )\rangle 
&= g_{1}(q^{2}) \Bigl[ p\cdot q \; \epsilon \cdot \varepsilon^{\ast} \notag \\
&- \, p\cdot \varepsilon^{\ast} \, q\cdot \epsilon \Bigr], 
\end{align}%
here \(m_{D^{*-}}\), \(m_{D^{+}}\), \(m_{\chi_{c1}}\), \(f_{D^{*-}}\), \(f_{D^{+}}\),  \(f_{\chi_{c1}}\) are the masses and decay constants of the mesons \(D^{*-}\), \(D^{+}\) and \( \chi_{c1}\),  respectively, where we denote the strong coupling for $ \chi_{c1} \rightarrow D^{*-} D^{+} $ by $g_{1}(q^2)$.%

The expression for the strong coupling constant $g_1(q^2)$ depends on the spectroscopic parameters of the molecular state $\chi_{c1}$, as well as on the masses and decay constants of the $D^{*\pm}$ and $D^{\mp}$ mesons, which are used as input parameters in the numerical calculations.
These parameters are listed in Table~\ref{tab:nput params2}.
Using the hadronic matrix elements, we construct the correlation function $\Pi^{\mathrm{Phys}}(p, p')$ in terms of hadronic parameters; the phenomenological part takes the following form:%
\begin{align}
\Pi _{1\mu \nu }^{\mathrm{Phys}}(p,p^{\prime }) &= g_{1}(q^{2})\frac{ f_{\chi_{c1}} m_{\chi_{c1}} f_{D^{+}} m_{D^{+}}^{2} f_{D^{*-}} m_{D^{*-}}}{(m_{c}+m_{d})\left(p^{2}-m^{2}\right) \left( p^{\prime 2}-m_{D^{*-}}^{2}\right) } \notag \\
&\times \frac{1}{q^{2}-m_{D^{+}}^{2}} \Bigg[ \frac{m^{2}-m_{D^{*-}}^{2}+q^{2}}{2}g_{\mu \nu } - p_{\mu }p_{\nu } \notag \\
&+ p_{\mu }^{\prime }p_{\nu } - \frac{m^{2}}{m_{D^{*-}}^{2}} p_{\mu }^{\prime }p_{\nu }^{\prime } + \frac{m^{2}+m_{D^{*-}}^{2}-q^{2}}{2m_{D^{*-}}^{2}} \notag \\
&\times p_{\mu }p_{\nu }^{\prime } \Bigg] + \cdots . \label{eq:phenomenological part }
\end{align}%

Applying the double Borel transformation with respect to $p^2$ and $p'^2$
will efficiently suppress the higher states and the continuum, leading to better convergence of the sum rules.%
\begin{table}[h!]
\centering
\caption{Masses and decay constants of the $D^{*\pm}$ and $D^{\pm}$ mesons used as input parameters in the numerical analysis.}
\label{tab:nput params2}
\begin{tabular}{|l|l|}
\hline
\textbf{Parameter} & \textbf{Value ($\mathrm{MeV}$)}  \\ \hline
$m_{D^{*\pm}}$ & $2010.26\pm 0.05$ \cite{ParticleDataGroup20245} \\\hline
$m_{D^{\pm}}$ & $1869.65\pm 0.05$ \cite{ParticleDataGroup20245} \\\hline
$f_{D^{\pm}}$ & $ 212.7\pm 0.6$  \cite{Bazavov:2018omf}\\\hline
$f_{D^{*\pm}}$ & $ 263\pm21$ \cite{Wang:2015mxa}\\ \hline
\end{tabular}
\end{table}%

To calculate the QCD side of the three-point sum rule, we write the correlation function $\Pi(p, p')$ in terms of the heavy- and light-quark propagators, and obtain the following expression: %
\begin{eqnarray}\label{eq:CF3}
\Pi_{1\mu\nu}^{\mathrm{OPE}}(p,p^{\prime}) &=&- \frac{1}{\sqrt{2}} \int d^{4}x \, d^{4}y \, e^{ip^{\prime} \cdot y} e^{-ip \cdot x} \nonumber    \\
& \times & \Bigg\{ \mathrm{Tr}\left[ \gamma_{\mu} S_{d}^{ia}(y-x) \, \gamma_{\nu} S_{c}^{ai}(x-y) \right] \nonumber \\
& \times & \mathrm{Tr}\left[ \gamma_{5} S_{c}^{jb}(-x) \, \gamma_{5} S_{d}^{bj}(x) \right] \nonumber \\
& - & \mathrm{Tr}\left[ \gamma_{\mu} S_{d}^{ib}(y-x) \, \gamma_{5} S_{c}^{bi}(x-y) \right] \nonumber \\
& \times & \mathrm{Tr}\left[ \gamma_{5} S_{c}^{ja}(-x) \, \gamma_{\nu} S_{d}^{aj}(x) \right] \Bigg\}. 
\end{eqnarray}%

The correlator $\Pi^{\mathrm{OPE}}(p,p^{\prime})$ is calculated by taking into account the nonperturbative contributions up to dimension $5$ and, similarly to $\Pi^{\mathrm{Phys}}(p,p^{\prime})$ it contains the Lorentz structure in the form of the term $\mathrm{g_{\mu\nu}}$.
It is shown that the contributions from operators of dimension higher than 5 are small and can easily be neglected without significantly changing the results. 
With the relevant invariant amplitude denoted by $\Pi^{\mathrm{OPE}}(p^{2},p^{\prime 2},q^{2})$, we match its double Borel transformation $\mathcal{B}\Pi^{\mathrm{OPE}}(p^{2},p^{\prime 2},q^{2})$ with that of $\mathcal{B}\Pi^{\mathrm{Phys}}(p^{2},p^{\prime 2},q^{2})$, then we apply continuum subtraction and obtain the sum rule for the strong coupling $g_{1}(q^{2})$.

After carrying out the double Borel transformation and subtracting the continuum, the QCD side takes the following form:%
\begin{eqnarray}
\Pi_{1} (\mathbf{M}^{2},\mathbf{s}_{0},q^{2})&=&%
\int_{4(m_{c}+m_{d})^{2}}^{s_{0}}ds\int_{(m_{c}+m_{d})^{2}}^{s_{0}^{\prime
}}ds^{\prime }\rho_{1} (s,s^{\prime },q^{2})  \notag \\
&\times & e^{-s/M_{1}^{2}}e^{-s^{\prime }/M_{2}^{2}},  \label{eq:QCD side}
\end{eqnarray}%
where $\rho_{1}(s,s',q^2)$ is the spectral density and $M^2 = (M_1^2,M_2^2)$ and $\mathbf{s}_0 = (s_0,s_0')$ are the Borel and continuum threshold parameters. The pairs $(M_{1}^{2}, s_{0})$ and $(M_{2}^{2}, s_{0}^{\prime})$ correspond to the channels of the molecule $\chi_{c1}$ and the $D^{*-}$ meson, respectively.
Matching both sides of the sum rule, the strong coupling $g_1(q^2)$ is extracted as in Eq.~(\ref{g1}). Hereafter we abbreviate $m \equiv m_{\chi_{c1}}$ and
$f \equiv f_{\chi_{c1}}$:
\begin{eqnarray}\label{g1}
g_{1}(q^{2}) &=& \frac{2(m_{c}+m_{d})}{f m f_{D^{+}} m_{D^{+}}^{2} f_{D^{*-}} m_{D^{*-}}} \; \frac{q^{2} - m_{D^{+}}^{2}}{m^{2} - m_{D^{*-}}^{2} + q^{2}} \notag \\
& \times & e^{m^{2}/M_{1}^{2}} \; e^{m_{D^{*-}}^{2}/M_{2}^{2}} \; \Pi_{1}(\mathbf{M}^{2}, \mathbf{s}_{0}, q^{2}).
\end{eqnarray}%
Given that the sum rules are assumed reliable solely within the spacelike region ($Q^2=-q^2>0$), we compute $g_1(Q^2)$ at fixed $Q^2$ values over the range $1\text{--}5~\mathrm{GeV}^2$.%

A further step for the numerical analysis consists of determining the working intervals for the Borel and continuum subtraction parameters, namely, $M^{2}$ and $s_{0}$. The bounds on $M^{2}$ and $s_{0}$ are the usual ones in the sum rule analysis, and they were discussed in the previous section. The intervals for $M_{1}^{2}$ and $s_{0}$ corresponding to the $ \chi_{c1} \rightarrow D^{*-} D^{+} $ channel are determined by Eq. (\ref{eq:Wind1}). The parameters $(M_{2}^{2},\ s_{0}^{\prime })$ for the channel of the $D^{*-}$ meson are changed within bounds 
\begin{equation}
M_{2}^{2}\in \lbrack 2.5,3.5]~\mathrm{GeV}^{2},\ s_{0}^{\prime }\in \lbrack
5.7,5.8]~\mathrm{GeV}^{2}.  \label{eq:Wind3}
\end{equation}%
Regions of the $M^{2}$ and $s_0$ are chosen to suppress their contributions to the coupling\, $g_{1}(Q^{2})$. An appropriate analytic function is used to fit the extracted values and extrapolate to the physical plane $Q^2 = -m_{D^{+}}^2$:%
\begin{equation}
\mathcal{G}_i(Q^2) = \mathcal{G}_i^0 
\exp\left( c_i^1 \frac{Q^2}{m^2} + c_i^2 \left( \frac{Q^2}{m^2} \right)^2 \right). \label{eq:FitF1}
\end{equation}%
The parameters $\mathcal{G}_i^0$, $c_i^1$, $c_i^2$ are obtained from a fit to the QCDSR results.
Numerical calculations indicate that $\mathcal{G}_1^0 = 0.626 ~\mathrm{GeV}^{-1}$, $c_1^1 =0.032$, and $c_1^2 =-0.00134 $ lead to a good fit to the sum rule results.

The strong coupling at the physical point is finally given by:%
\begin{equation}
g_1 \equiv \mathcal{G}_1(-m_{D^{+}}^2). \label{eq:strong coupling first}
\end{equation}%
The strong coupling constants of the heavy-light mesons may be useful in understanding the final-state interactions in the heavy quarkonium (or meson) decays \cite{OsorioRodrigues:2013xvc}. Calculation of this strong coupling
constant in the space-like region and subsequent extrapolation to $Q^2=-m^2$, with m the mass of the off-shell meson, determines the coupling constant. 

\begin{widetext}

\begin{figure}[h!]
\begin{center}
\includegraphics[totalheight=6cm,width=8cm]{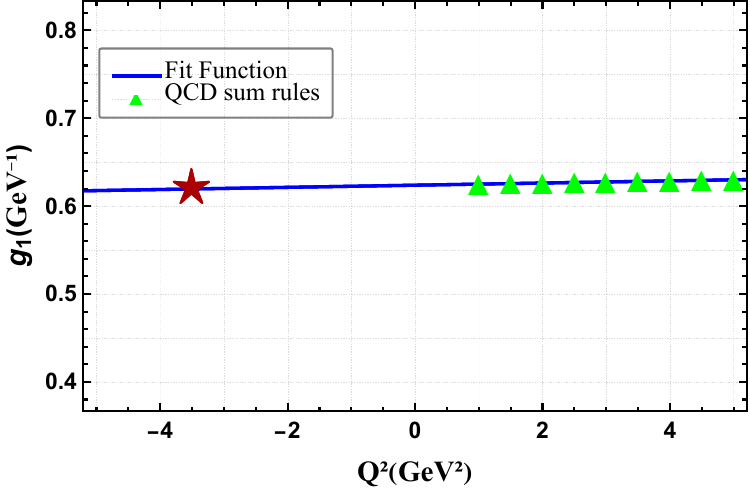}
\includegraphics[totalheight=6cm,width=8cm]{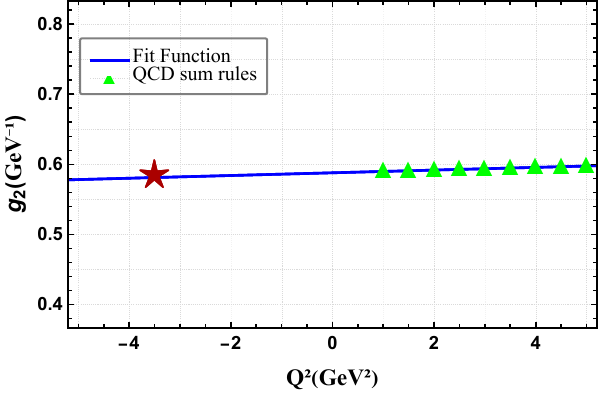}
\end{center}
\caption{The sum-rule predictions together with the fitted curve for the strong coupling $g_{1,2}(Q^{2})$ are shown. The red star marks the physical point $Q^{2} = -m^{2}_{D^{\pm}}$.}
 \label{fig:g1 and g2}
\end{figure}

\end{widetext}

To reduce uncertainties from the extrapolation procedure, we perform the calculation twice, with each meson off-shell, usually the ones with the largest mass differences, yielding two different strong coupling constants. These functions must then take the same values at their poles, so we obtain a single coupling constant for the vertex. The calculation of the coupling constant is a primary result, whereas the physically strong coupling constants are derived quantities with values determined in experimental high-energy physics \cite{OsorioRodrigues:2013xvc, Yu:2015xwa}. This can be done within the QCDSR technique, provided that the sum rule is sufficiently stable \cite{Bracco:2011pg, CerqueiraJr:2021afn} as illustrated in Fig. \ref{fig:g1 and g2}. 

At the mass shell $q^2 = m_{D^{+}}^2$, this function  is equal
to %
\begin{equation}\label{eq:g1num}
g_1  \equiv \mathcal{G}_1(-m_{D^{+}}^2) = (0.62 \pm 0.08  )\
~\mathrm{GeV}^{-1},
\end{equation}%
and the corresponding partial width is computed as:
\begin{equation}
\Gamma \left[\chi_{c1} \rightarrow D^{*-} D^{+}\right]
=g_{1}^{2}\frac{\lambda _{1}}{24\pi m^2_{\chi_{c1}} }|M_{1}|^{2},
\end{equation}%
where%
\begin{eqnarray}
| M_{1}|^{2} &=& \frac{1}{4m_{D^{*-}}^{2}} \left[ m^{6} - 2m^{4}m_{D^{+}}^{2} + 2 \left( m_{D^{*-}}^{3} - m_{D^{+}}^{2} \right. \right. \notag \\
& \times & \left. \left. m_{D^{*-}} \right)^{2} + m^{2} \left( m_{D^{+}}^{4} + 6m_{D^{+}}^{2}m_{D^{*-}}^{2}-3m_{D^{*-}}^{4} \right) \right] . \notag \\ 
\label{eq:M1 first channel}
\end{eqnarray}
Here, $\lambda_1 = \lambda(m, m_{D^+}, m_{D^{*-}})$ is the K\"all\'en 
(triangle) function, which gives the momentum of the final-state 
particles in the rest frame of the decaying particle. With $x$ the mass 
of the decaying particle and $y,z$ the masses of the two final-state 
particles,
\begin{equation}
\lambda(x,y,z) = \frac{\sqrt{x^4+y^4+z^4-2(x^2y^2+x^2z^2+y^2z^2)}}{2x}.
\label{eq:lambda}
\end{equation}
The coupling constant obtained above in Eq.~(\ref{eq:g1num}) allows for a straightforward calculation of the decay width for the process $\chi_{c1}\rightarrow D^{*-}D^{+}$: 
\begin{equation}
\Gamma[{\chi_{c1}} \rightarrow D^{*-} D^{+} ]= (30.53\pm 7.30)~\mathrm{MeV}.  \label{eq:DW first}
\end{equation}
\subsection{$\chi_{c1} \rightarrow D^{*+} D^{-}$}

The second decay channel, $\chi_{c1} \rightarrow D^{*+} D^{-}$, may be analyzed in a manner analogous to the initial process. Nonetheless, in this scenario, we utilize distinct interpolating currents that are aligned with the quantum numbers corresponding to the final-state mesons $D^{-}$ and $D^{*+}$. In particular, the following formulations for the interpolating currents are employed:%
\begin{align}
J_{\mu}^{D^{*+}}(x) &= \bar{d}_j(x) i \gamma_{\mu} c_j(x), \\
J^{D^-}(x) &= \bar{c}_i(x) i \gamma_5 d_i(x), \label{eq:current D}
\end{align}
where the indices $i$ and $j$ are color indices. 

To proceed with the sum rule analysis, the relevant matrix elements for $D^{-}$ and $D^{*+}$ mesons need to be defined, as will be discussed in the following steps. These matrix elements are essential to form the phenomenological side of the correlation function connected to the $\chi_{c1} D^{*+} D^{-}$ vertex
\begin{equation}
\langle 0 | J_{\mu}^{D^{*+}} | D^{*+} (p^{\prime},\varepsilon) \rangle = f_{D^{*+}} m_{D^{*+}} \varepsilon _{\mu }(p^{\prime }), \label{eq:matrrix D positive}
\end{equation}
\begin{equation}
\langle 0 | J^{D^{-}} | D^{-}(q) \rangle = \frac{f_{D^{-}} m_{D^{-}}^2 }{m_c + m_d}. \label{eq:Matrix D negative elemen}
\end{equation}%

We express the vertex $\chi_{c1} D^{*+} D^{-}$  by the matrix element 
\begin{eqnarray}
\langle D^{-}(q)D^{*+} (p^{\prime },\varepsilon )|\chi_{c1}(p,\epsilon )\rangle 
&=& g_{2}(q^{2}) \Bigl[ p\cdot q \; \epsilon \cdot \varepsilon^{\ast} \notag \\
& - & p\cdot \varepsilon^{\ast} \, q\cdot \epsilon \Bigr].
\end{eqnarray}
 The strong coupling associated with the process $\chi_{c1} \rightarrow D^{*+} D^{-} $ is denoted as $g_{2}(q^2)$. Employing these expressions, the novel representation of the correlation function $ \Pi ^{\mathrm{Phys}}(p, p') $ can be readily derived, leading to the phenomenological side as follows:%
\begin{eqnarray}
&&\Pi _{2\mu \nu }^{\mathrm{Phys}}(p,p^{\prime })=g_{2}(q^{2})\frac{f m 
f_{D^{-}}m_{D^{-}}^{2}f_{D^{*+} }m_{D^{*+} }}{(m_{c}+m_{d})\left(
p^{2}-m^{2}\right) \left( p^{\prime 2}-m_{D^{*+} }^{2}\right) }  \notag \\
&&\times \frac{1}{(q^{2}-m_{D^{-}}^{2})}\left[ \frac{m^{2}-m_{D^{*+}
}^{2}+q^{2}}{2}g_{\mu \nu }-p_{\mu }p_{\nu }\right.  \notag \\
&&\left. +p_{\mu }^{\prime }p_{\nu }-\frac{m^{2}}{m_{D^{*+} }^{2}}p_{\mu
}^{\prime }p_{\nu }^{\prime }+\frac{m^{2}+m_{D^{*+} }^{2}-q^{2}}{%
2m_{D^{*+} }^{2}}p_{\mu }p_{\nu }^{\prime }\right]  \notag \\
&&+\cdots .  \label{eq:phenomenological second part}
\end{eqnarray}%

Applying a double Borel transformation to both $p^2$ and $p'^2$ suppresses contributions from higher states and the continuum, leading to improved convergence of the sum rules. To be able to study the QCD side of the three-point sum rule, we write $\Pi(p,p^{\prime})$ in terms of quark propagators as follows:
\begin{eqnarray}\label{eq:CF4}
\Pi_{2\mu\nu}^{\mathrm{OPE}}(p,p^{\prime}) &=&- \frac{1}{\sqrt{2}} \int d^{4}x \, d^{4}y \, e^{ip^{\prime} \cdot y} e^{-ip \cdot x} \nonumber \\
&\times & \Bigg\{ \mathrm{Tr}\left[ \gamma_{\mu} S_{c}^{ib}(y-x) \, \gamma_{5} S_{d}^{bi}(x-y) \right] \nonumber \\
& \times  &\mathrm{Tr}\left[ \gamma_{5} S_{d}^{ja}(-x) \, \gamma_{\nu} S_{c}^{aj}(x) \right] \nonumber \\
& - & \mathrm{Tr}\left[ \gamma_{\mu} S_{c}^{ia}(y-x) \, \gamma_{\nu} S_{d}^{ai}(x-y) \right] \nonumber \\
& \times  &\mathrm{Tr}\left[ \gamma_{5} S_{d}^{jb}(-x) \, \gamma_{5} S_{c}^{bj}(x) \right] \Bigg\}.
\end{eqnarray}%

The correlator $\Pi^{\mathrm{OPE}}(p,p^{\prime})$ is found using nonperturbative contributions up to dimension $5$, and like the correlator $\Pi^{\mathrm{Phys}}(p,p^{\prime})$, it contains the Lorentz structure proportional to the term $\mathrm{g_{\mu\nu}}$.
Our analysis demonstrates that the contributions of operators with dimensions higher than 5 are sufficiently suppressed and can therefore be neglected without introducing any appreciable changes to the final results.
We denote the appropriate invariant amplitude by $\Pi^{\mathrm{OPE}}(p^{2},p^{\prime 2},q^{2})$, and we match its double Borel transformation $\mathcal{B}\Pi^{\mathrm{OPE}}(p^{2},p^{\prime 2},q^{2})$ to the Borel-transformed and continuum-subtracted phenomenological side $\mathcal{B}\Pi^{\mathrm{Phys}}(p^{2},p^{\prime 2},q^{2})$, and again after performing the continuum subtraction, we obtain the sum rule for the coupling $g_{2}(q^{2})$.%

After double Borel transformation and continuum subtraction, the QCD side can be written as: 
\begin{eqnarray}
\Pi_{2} (\mathbf{M}^{2},\mathbf{s}_{0},q^{2})&=&%
\int_{4(m_{c}+m_{d})^{2}}^{s_{0}}ds\int_{(m_{c}+m_{d})^{2}}^{s_{0}^{\prime
}}ds^{\prime }\rho_{2} (s,s^{\prime },q^{2})  \notag \\
&\times & e^{-s/M_{1}^{2}}e^{-s^{\prime }/M_{2}^{2}},  \label{eq:QCD second}
\end{eqnarray}%
where $\rho_{2}(s,s',q^2)$ is the spectral density and $M^2 = (M_1^2, M_2^2)$ and $\mathbf{s}_0=(s_0,s_0')$ denote the Borel and continuum threshold parameters.
The pairs $(M_{1}^{2}, s_{0})$ and $(M_{2}^{2}, s_{0}^{\prime})$ correspond to the channels of the molecule $\chi_{c1}$ and the $D^{*+}$ meson, respectively. 
Matching both sides of the sum rule, we obtain the strong coupling $g_2(q^2)$ as:
\begin{eqnarray}\label{g2}
g_{2}(q^{2}) &=& \frac{2(m_{c}+m_{d})}{f m f_{D^{-}} m_{D^{-}}^{2} f_{D^{*+}} m_{D^{*+}}} \; \frac{q^{2} - m_{D^{-}}^{2}}{m^{2} - m_{D^{*+}}^{2} + q^{2}} \notag \\
& \times & e^{m^{2}/M_{1}^{2}} \; e^{m_{D^{*+}}^{2}/M_{2}^{2}} \; \Pi_{2}(\mathbf{M}^{2}, \mathbf{s}_{0}, q^{2}).
\end{eqnarray}%

Since the sum rules are reliable only in the spacelike region 
($Q^2 = -q^2 > 0$), we compute $g_2(Q^2)$ at fixed values of $Q^2$ 
over the range $1$--$5~\mathrm{GeV}^2$.
Employing the constraints from Eqs.\ (\ref{eq:Wind1}) and (\ref{eq:Wind3}), we determine the working windows of the Borel parameter $M^2$ and the continuum threshold $s_{0}$. The resultant values are fitted using an appropriate analytic function that facilitates extrapolation to the physical region $Q^2 = -m_{D^{-}}^2$:%

\begin{equation}
\mathcal{G}_i(Q^2) = \mathcal{G}_i^0 
\exp\left( c_i^1 \frac{Q^2}{m^2} + c_i^2 \left( \frac{Q^2}{m^2} \right)^2 \right). \label{eq:FitF2}
\end{equation}%
The parameters, $\mathcal{G}_i^0$, $c_i^1$ and $c_i^2$ are derived from a fit to the QCDSR results. Numerical calculations indicate that $\mathcal{G}_2^0 =0.592\mathrm{GeV}^{-1}$, $c_2^1 =0.053$ and $c_2^2 = -0.0035$ lead to a very good agreement with the sum rule predictions. 
At the end, the strong coupling at the physical point is:%
\begin{equation}
g_2  \equiv \mathcal{G}_2(-m_{D^{-}}^2), \label{eq:Coupl str 2}
\end{equation}%
As a result, at the mass shell $q^2 = m_{D^{-}}^2$, this function takes the value %

\begin{equation}\label{eq:g2num}
g_2  \equiv \mathcal{G}_2(-m_{D^{-}}^2) =  (0.59 \pm 0.07)\
~\mathrm{GeV}^{-1}.
\end{equation}%
and the corresponding partial width takes the form:
\begin{equation}
\Gamma \left[\chi_{c1} \rightarrow D^{*+} D^{-}\right]
=g_{2}^{2}\frac{\lambda _{2}}{24\pi m^2_{\chi_{c1}}}|M_{2}|^{2},
\end{equation}%
with $\lambda_{2}=\lambda (m, m_{D^-}, m_{D^{*+}})$ taken from Eq. (\ref{eq:lambda}), where%

\begin{eqnarray}
|M_{2}|^{2} &=& \frac{1}{4m_{D^{*+}}^{2}} \left[ m^{6} - 2m^{4}m_{D^{-}}^{2} + 2 \left( m_{D^{*+}}^{3} - m_{D^{-}}^{2} \right. \right. \notag \\
& \times & \left. \left. m_{D^{*+}} \right)^{2} + m^{2} \left( m_{D^{-}}^{4} + 6m_{D^{-}}^{2}m_{D^{*+}}^{2} - 3m_{D^{*+}}^{4} \right) \right] . \notag \\ 
\label{eq:M2 second decay width}
\end{eqnarray}%

The coupling constant obtained from Eq.~(\ref{eq:g2num}) allows for a direct calculation of the decay width of the channel ${\chi_{c1}} \rightarrow D^{*+} D^{-}$: %
\begin{equation}
\Gamma[\chi_{c1} \rightarrow D^{*+} D^{-}] = (27.11 \pm 6.30 )~\mathrm{MeV}.  \label{eq:decay width second}
\end{equation}%

Using data concerning the partial decay widths of the principal channels associated with the hadronic molecule  $\chi_{c1}  D^{*\pm} D^{\mp}$, one can readily approximate its total decay width%
\begin{equation}
\Gamma[\chi_{c1} \rightarrow D^{*\pm} D^{\mp}] = (57.64 \pm 9.64 )~\mathrm{MeV},  \label{eq:DW4}
\end{equation}%
which indicates that $\chi_{c1}$ constitutes a relatively broad resonance. Within theoretical and experimental uncertainties, our prediction for the width of the hadronic molecule 
$\chi_{c1}$ is compatible with the LHCb value in Eq.~(\ref{eq:Data_chi}).%

\section{Conclusions}\label{sec:Conclusions}

We have explored the axial-vector state  $\chi_{c1}$ as a hadronic molecule using the QCDSR approach. The predicted mass, $m_{\chi_{c1}}=(4005.89 \pm 47.15)~\text{MeV}$, agrees with the LHCb measurement~\cite{LHCb:2024vfz}, consistent with a molecular interpretation of this state. Although the state is unstable against strong dissociation into $D^{*\pm}D^{\mp}$, we calculated the corresponding partial decay widths using the three-point QCD sum rule technique, with quark-hadron duality used to constrain the strong couplings at the relevant meson-meson vertices. 
The resulting total width, $\Gamma_{\chi_{c1}} = (57.64 \pm 9.64)~\text{MeV}$, is consistent with the LHCb result as well.

The implications of $\chi_{c1}(4010)$ for the nature of $X(3872)$ are significant. Within the $\bar{D}D^{*}$ scattering framework, both states emerge simultaneously as poles of the scattering amplitude, forming a hadronic-molecular/charmonium mixed doublet straddling the $\bar{D}D^{*}$ threshold \cite{Li:2024pfg, Gao:2025tob, Shi:2024llv}. The agreement between our derived parameters and the LHCb result \cite{LHCb:2024vfz} supports the molecular interpretation of  $\chi_{c1}$ \cite{Miyake:2025ktz, Miyake:2025mmp}. 

The coupling constants extracted in this work are important inputs for describing hadronic dynamics, including the treatment of final-state interactions in heavy quarkonium decay and the calculation of absorption cross sections relevant to heavy quarkonium suppression in hadronic matter. While the present results suggest $\chi_{c1}$ as a viable axial-vector molecular state, a fully conclusive picture of its internal structure would benefit from a complementary study of its spectroscopic properties within a diquark-antidiquark (compact tetraquark) configuration, allowing a direct comparison between the two pictures alongside further experimental data. We hope that upcoming high-precision measurements at BESIII, LHCb, Belle~II, CEPC, FCC, and ILC will provide further experimental constraints and deepen our understanding of the dynamics of this exotic state.%
\begin{widetext}
\section*{DECLARATION OF AI-ASSISTED LANGUAGE EDITING}
AI assistance was limited to language editing and improving readability. The authors reviewed and revised all such edits and remain fully responsible for the manuscript's scientific content.
\end{widetext}
\bibliographystyle{unsrt}

\end{document}